\documentclass{mem}
\usepackage{natbib}\usepackage{txfonts}\usepackage{balance}
\usepackage{graphicx}
\usepackage[a4paper]{hyperref}
\idline{75}{282}
\begin{document}
\def\teff{$T\rm_{eff }$}
\def\kms{$\mathrm {km s}^{-1}$}
\def\ms{$\mathrm {m\,s}^{-1}$}
\def\daa{$\delta\alpha/\alpha\,$}
\def\zabs{$z\rm_{abs}\,$}
 \def\zem{$z\rm_{em}\,$}
 \def\hh{H$_2\,\,$}

\title{Calibration issues in  estimating variability of the fine structure constant (alpha) with cosmic time}

\subtitle{}

\author{
Miriam \,Centuri\'on\inst{1} 
Paolo Molaro\inst{1}
\and Sergei Levshakov\inst{2}
}

  \offprints{M. Centuri\'on}

\institute{
Istituto Nazionale di Astrofisica --
Osservatorio Astronomico di Trieste, Via Tiepolo 11,
I-34131 Trieste, Italy \\
\email{centurion@oats.inaf.it} 
\and
Ioffe Institute Politekhnicheskaya Str. 26, 194021St. Petersburg, 
Russia 
}

\authorrunning{Centuri\'on}

\titlerunning{Calibration issues in \daa }

\abstract{ Laser Comb Wavelength calibration  shows that the ThAr one  is locally unreliable with possible deviations of up to 100 \ms  within one order range, while delivering an overall 1 \ms accuracy (Wilken et al 2009). Such deviation corresponds to \daa   $\approx 7\cdot 10^{-6}$ for a Fe\,{\sc ii}-Mg\,{\sc ii} pair. Comparison of line shifts  among the 5 Fe\,{\sc ii} lines, with almost identical sensitivity to fine structure constant  changes, offers a  clean way to   directly test  the presence of possible local  wavelength calibration errors of whatever origin.  We analyzed  5 absorption systems, with \zabs ranging from 1.15 to 2.19 towards 3 bright QSOs. The results show that  while some lines are aligned within 20 \ms,  others  reveal    large  deviations reaching 200 \ms or higher  and corresponding to a \daa  $\ge 10^{-5}$ level. The origin of these deviations is not clearly  identified but  could be related to the adaptation of wavelength calibration to CCD manufacturing irregularities. These results  suggest that to draw  conclusions from  \daa analysis based on  one or only few lines must be done with extreme care. 
 
\keywords{Atomic data -- line:profiles -- methods: data analysis -- techniques: spectroscopic -- quasars: absorption lines}
}
\maketitle

\section{Introduction}
Investigations into possible systematic effects entering  in the measurement of \daa have been discussed  among others in  Murphy et al (2003), Chand et al (2006), Molaro et al (2008).  Here we focus onto the wavelength calibration error.

The wavelength calibration of the QSO CCD images is made  by comparison with ThAr lamp taken before and after  the QSO frames.
The laboratory wavelengths used in the line identification have uncertainties  between 10-150 \ms (Palmer \& Engleman 1983, Lovis \& Pepe 2007 ).  Errors in the wavelength calibration could be  estimated from the mean wavelength-pixel residuals from the polynomial best solution. Residuals vary   from   3-4 m\AA   (Chand et al 2006) to  1 m\AA, or even better,     (Levshakov et al 2006, Thompson et al 2009). Murphy et al (2003) treating the ThAr as if they were QSO lines obtained  that the error in the individual absorption systems is typically few times $10^{-6}$ but  there are significant cases up to  $5 \cdot 10^{-6}$  or higher.
When averaged over a  large sample of  simulated systems they obtained  \daa = $ 0.4 \pm 0.8\cdot 10^{-7}$ and concluded that these errors  are not  expected  to drive line shifts in the QSO absorption-line results.  A similar conclusion was  achieved by Chand et al (2006) where  3 and 2 ThAr lines were  taken in proximity of FeII and MgII lines, respectively.

 However, the residuals in the wavelength calibration may not reflect  the real wavelength error entirely.
Recently an experiment conducted at HARPS with the first Laser Comb  wavelength calibration  in the optical revealed  significant deviations of the ThAr wavelength calibration  with  peak-to-peak deviations of up to 100 \ms within one echelle order.
 Similarly in HIRES-Keck observations  a comparison of the ThAr calibration with a I2 self-calibration spectrum revealed significant and not reproducible deviations (Griest et al 2009)

\section{The method and data analysis}

The spectroscopic measurability of \daa is based on the fact that the
energy of each line transition depends  on 
$\alpha$.  The value of \daa\  follows from the 
measured relative radial velocity shifts, $\Delta$v, between
lines with different sensitivity coefficients.
\begin{equation}
\frac{\delta\alpha}{\alpha} = \frac{(\mathrm v_2 - \mathrm v_1)}
{2\,c\,({\cal Q}_1 - {\cal Q}_2)} = 
\frac{\Delta \mathrm v}{2\,c\,\Delta {\cal Q}}\ .
\label{E1}
\end{equation} 
Where  ${\cal Q} = q/\omega_0$
Dzuba et al. (2002) and  
Porsev et al. (2007). The Fe\,{\sc ii} $\lambda\lambda$ 2600.1722,  2586.6494,  2382.7641, 2374. 4601 and  2344.2128  lines have  similar sensitivity to \daa,  so they should be always found aligned. The measurement  of their relative shifts are a clear tool to monitor the quality of the wavelength calibration.

\begin{figure*}[]
\resizebox{\hsize}{!}{\includegraphics[clip=true]{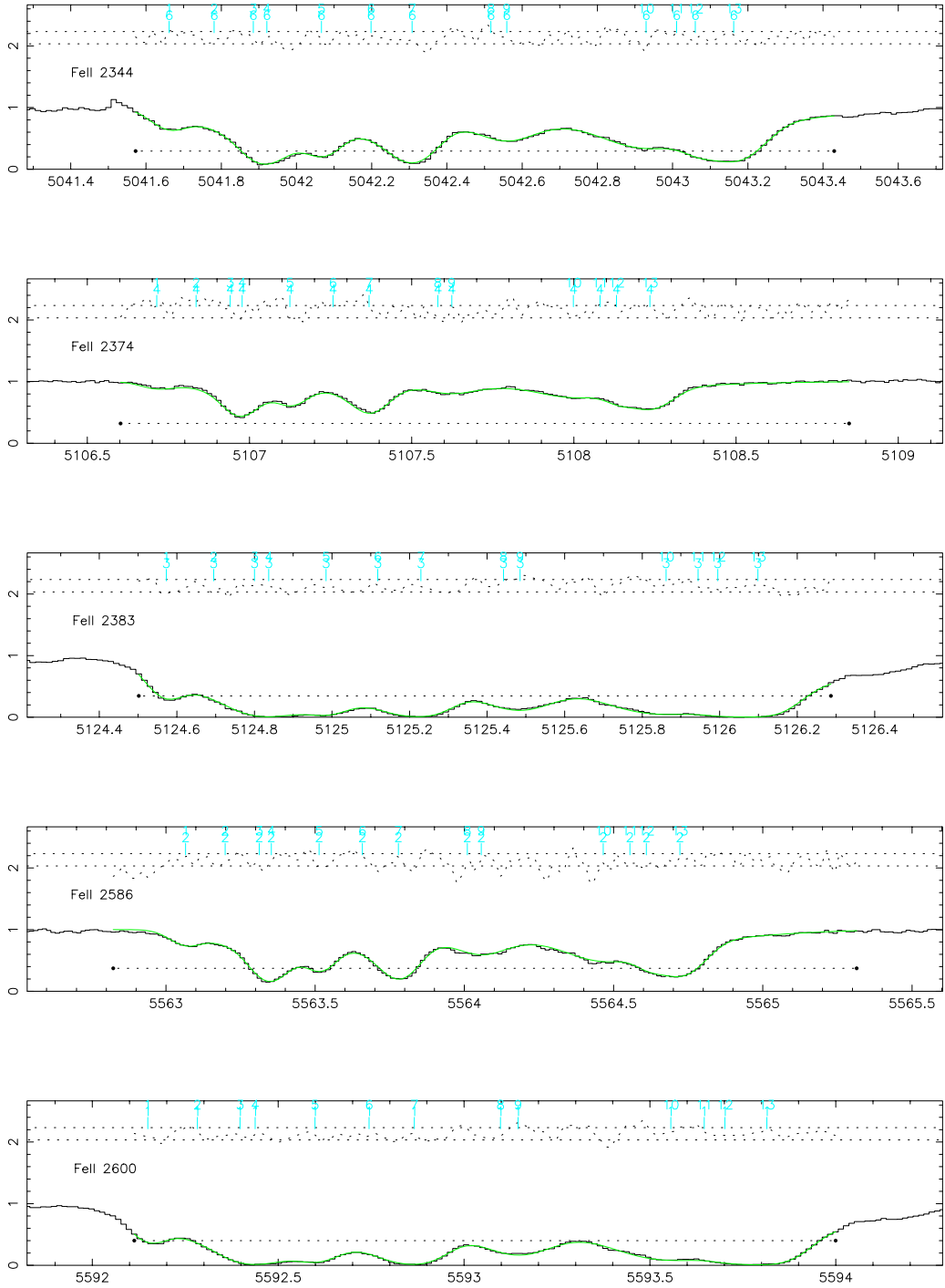}}
\caption{\footnotesize
The 5 FeII lines of the \zabs =1.15 in HE 0515 -- 4414 are shown. The fit is obtained with VPFIT by using  13  components ($\chi^2$=1.1). Treating the Fe\,{\sc ii} $\lambda\lambda$2383 as Mg\,{\sc ii} $\lambda\lambda 2804$ the  whole set of lines would provide a spurious signal at  \daa = $( -11.9 \pm 2.5) \cdot 10^{-6}$. }
\end{figure*}

For the present analysis we used 5  systems towards  QSOs HE 0515--4414, QSO 1101--264 and HE 0001--2340.
 The wavelength calibration was done  with the most recent recipes except for  HE 0001--2340, for which we used the same data  of Chand et al (2004) and Murphy et al (2008)   to allow a direct comparison.
 
 To measure relative velocity shifts we used VPFIT and  assigned to  Fe\,{\sc ii} $\lambda\lambda$ 2383 the $q$ factor of Mg\,{\sc ii} $\lambda\lambda 2804$   acting  as  the reference line.   The comparison between this line and any of the 
Fe\,{\sc ii} $\lambda\lambda  2344, 2374,  2586, 2600$ transitions has an effective $\Delta {\cal Q} \approx 0.035$.

\begin{figure*}[]
\resizebox{\hsize}{!}{\includegraphics[height=0.7\linewidth]{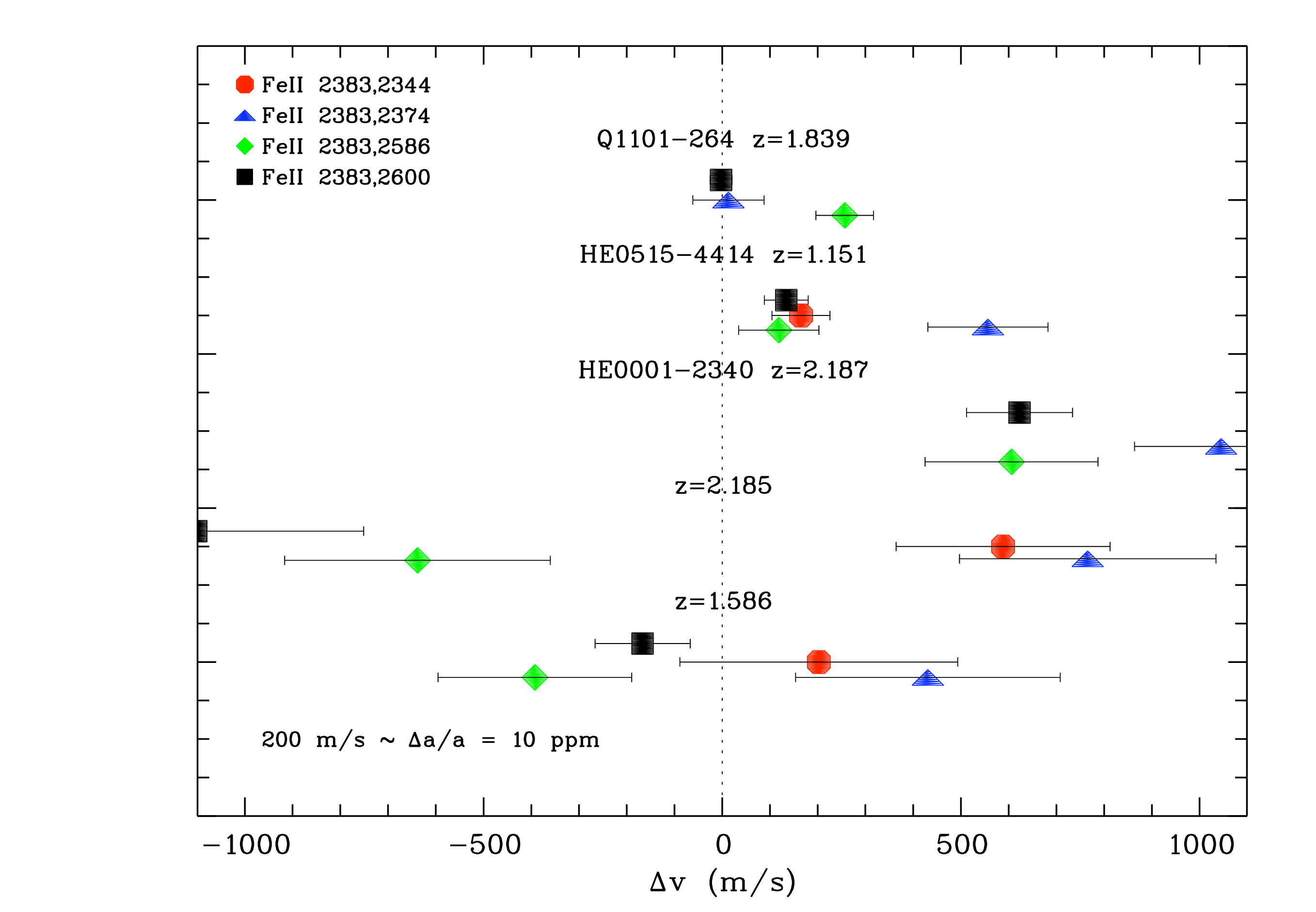}}
\caption{
\footnotesize
Velocity shifts between the 5 Fe\,{\sc ii} lines which share almost identical sensitivity  to
changes in the fine structure constant (ppm, part per million = 10$^{-6}$).  
}
\end{figure*}

\section{Results} 

\subsection {Q 1101-264}
 
  We found that two  Fe\,{\sc ii}  are perfectly aligned to our reference line  Fe\,{\sc ii} $\lambda\lambda$ 2383  
 with velocity shifts of $-2.5 \pm 23$ \ms and $13\pm 74.6$  \ms, respectively. Here the error stems from the photon noise  and the uncertainties in  modeling the absorption, while  $\Delta \mathrm v$ values reflect the lack of a real shift between the lines. On the other hand    Fe\,{\sc ii} 2586  strongly deviates from the other two,  
 $256.6 \pm 60.4$ \ms,  revealing an intrinsic wavelength shift between the  lines.
 Such a deviation  cannot be justified on the basis of the wavelength calibration residuals which are $\approx$  60 \ms. This line falls at the edge of the order and it may be possible that we are witnessing an effect similar to  that found by Wilken et al (2009) with HARPS.  
 
 \subsection {HE 0515-4414}
 
 We found 3 lines that  deviate  by  $\approx (135 \pm  50)$ \ms  with respect to Fe\,{\sc ii} 2383,  and the Fe\,{\sc ii} 2374  at  $(556 \pm 125.8)$ \ms showing  a strong an unjustified deviation from all the others lines. The latter would correspond  to  a spurious signal of  \daa = $(-26.1\pm 5.9)$ ppm.
 
 \subsection{ HE 0001-2340} 
 
Towards this QSO  we analyzed 3 absorption systems. They all are relatively simple but surprisingly we obtained a poor reduced $\chi^2$ in the  modeling. Correspondingly the error in the line positions  are quite large. The system at \zabs = 2.187  gives the most precise results  showing 3 lines clearly shifted with respect to our reference,  and the Fe\,{\sc ii} $\lambda\lambda$ 2374 further shifted away from the reference by  $(1044.9\pm 181.25$)  \ms. It may be possible that   a rather  old data reduction is also  affecting these data.

\section{Conclusions}

By using line shifts measurements of the  five Fe\,{\sc ii} lines with  identical sensitivity to $\alpha$ changes we have  studied   possible systematics in the wavelength calibration. We  found that:
 \begin{itemize}
 \item{In most of the systems one or more lines are deviating   by several hundreds of \ms. This exceeds the  likely errors estimated from the residuals  and reveals the presence of hidden systematics in the spectral data. }
 \item{ These errors could  severely affect the \daa measure,  in particular  in the methods where \daa is derived from few lines or where some lines are crucial for the result. To have a reliable measure it is important to have consistent \daa obtained from several lines for a given absorption system.
 }
 \item{HE 0001-2340, which is one of the QSO used in the present UVES controversy,  reveals very poor quality and the whole sample should be reanalyzed {\it ab initio} }
 \end{itemize}
  
\bibliographystyle{aa}

\end{document}